\documentclass[prd,twocolumn,showpacs,nofootinbib, aps,10pt,superscriptaddress,amsmath,amssymb]{revtex4-1}
\usepackage{graphicx}% Include figure files
\usepackage{dcolumn}% Align table columns on decimal point
\usepackage{bm}% bold math
\usepackage{epstopdf}
\usepackage{mathtools}
\usepackage{natbib}
\usepackage{captcont}
\usepackage{booktabs}
\usepackage{threeparttable}

\begin{document}
\title{Constraining primordial black holes in dark matter with kinematics of dwarf galaxies}
\author{Bo-Qiang Lu}\email{bqlu@itp.ac.cn}
\affiliation{State Key Laboratory of Theoretical Physics,
Institute of Theoretical Physics, Chinese Academy of Sciences, Beijing, 100190, China}
\affiliation{Department of Physics, National Taiwan University, Taipei, 10617, Taiwan, Republic of China}
\author{Yue-Liang Wu}\email{ylwu@itp.ac.cn}
\affiliation{State Key Laboratory of Theoretical Physics, 
Institute of Theoretical Physics, Chinese Academy of Sciences, Beijing, 100190, China}
\affiliation{University of Chinese Academy of Sciences, Beijing, 100049, China}

\begin{abstract}

We propose that the kinematical observations of dwarf galaxies can be used to constrain the primordial black hole's (PBH) 
abundance in dark matter since the presence of primordial black holes in star clusters will lead to the radial 
velocity dispersion of the system.
%Using the velocity dispersion observations from Leo I as an example we show that the primordial black hole fraction 
For instance, using the velocity dispersion observations from Leo I we show that the primordial black hole fraction 
$f_{\rm PBH}\gtrsim 2.0\times(1~M_{\odot}/m_{\rm PBH})^2$ is ruled out at a 99.99\% confidence level.
This method yields the most stringent limits on the PBH abundance at the mass scales $\sim (1-10^3)~M_{\odot}$
and tightly constrains the primordial origin of gravitational wave events observed by the LIGO experiments.
%and for the first time, we have the confidence to conclude that the 
%primordial origin of the gravitational wave events observed by LIGO has been severely constrained.

\end{abstract}
\maketitle

\section{Introduction}

Despite overwhelming evidence from cosmological and astrophysical observations that shows about 85\% of the matter in the Universe 
consists of dark matter (DM), the nature of DM still remains a mystery in science today.
By now various experiments have been designed to aim at the weakly interacting massive particles (WIMPs), which have masses and 
coupling strengths at the electroweak scale \cite{Lee1977, Hut1977}. No obvious evidence for WIMPs has been observed either 
in direct \cite{Pandax2017, LUX2017} or indirect DM detections \cite{Fermi2015, Lu2016PRD} so far, and stringent 
limits have been set on this DM hypothesis.

One intriguing alternative to the particle DM is that a population of primordial black holes (PBHs) formed  
in the early Universe \cite{Carr1974, Carr1975, Meszaros1974} may contribute to the DM abundance.
The widely accepted mechanism to produce PBHs is the direct gravitational collapse of the large primordial curvature fluctuations 
in the early Universe \cite{Hawking1971},
other mechanisms for the formation of PBHs can be found in Refs. \cite{Garriga2016, Deng2017, Hawking1989, Polnarev1991}.
The detection of PBHs not only provides us with useful information of the early Universe but also sheds 
light on the inflation models \cite{Sasaki2018}.

It was suggested in Ref. \cite{Bird2016} that the gravitational waves (GWs) produced from the merger of two PBHs 
with masses $\sim 30~M_{\odot}$ may have been detected by the Advanced Laser Interferometer Gravitational-Wave Observatory (LIGO) \cite{GW150914}. 
Moreover, the merger rate estimated from the GW event observations can be explained if PBHs constitute a 
small fraction of DM \cite{Sasaki2016}.

The excellent places to test this hypothesis are star clusters in dwarf galaxies \cite{Brandt2016}, 
where dark matter is believed to be the dominant mass component.
A star cluster can be treated as a collisionless system when the relaxation time is long compared to the age of the Universe \cite{Merritt2013}.
When the PBHs with mass $m_{\rm PBH}\gg m_{\rm s}$ (the stellar mass $m_{\rm s}\sim 1~M_{\odot}$) are present in clusters,
mass segregation \cite{Spitzer1969} leads to the PBHs concentrating in the core of the cluster, 
while the stars are displaced to the larger radii.
This effect is limited by the observed half-light radius \cite{Brandt2016} and surface density \cite{Koushiappas2017} of dwarf galaxies.

%We pay attention to the kinematical observations of the dwarf galaxy.
Numerical investigations of the star cluster evolution \cite{Giersz1994, Peuten2017, Spurzem1995, Takahashi1996, Takahashi1997}
have shown that two-body relaxation generates a high degree of velocity anisotropy in the outer parts 
of the cluster \cite{Spitzer1987} and the appearance of the anisotropy is closely 
related to the energy transport processes between the different mass components.
For the age of stellar evolution $t_{\rm age}\lesssim 5\tau_{\rm rh}$ ($\tau_{\rm rh}$ is the half-mass relaxation time), 
the light components obtain radial kinematical energies from the gravitational encounters
with the heavy components at the cluster center. Consequently, they move to the outer regimes of the cluster along the 
radial orbits, which leads to a radial velocity dispersion of the cluster.
During the later stage, as the light components on radial orbits reach the boundary of the cluster,
they will escape from the bound system more efficiently if they acquire more kinematical energies from the gravitational encounters.
With the depletion of the stars on radial orbits, the stars on tangential orbits become dominant, which leads to a tangential
velocity dispersion.

%Anisotropy has been observed in many globular clusters. 
The observations of the high values of center velocity dispersion in $\omega$ Cen and 47 Tuc have suggested the presence of 
heavy nonluminous remnants (stellar black holes or neutron stars) concentrated in the cores of these clusters \cite{Meylan1986AA, Mann2018}.
In Ref. \cite{Meylan1986AA}, the velocity dispersion profiles are calculated by solving the Jeans' equation, 
taking into account the mass distributions of stars and heavy remnants (see Fig. 9 of the reference).
Their calculations show that 7\% of total mass of $\omega$ Cen and 2\% of total mass of 47 Tuc should be in the heavy remnants 
(with masses $\sim 2~M_{\odot}$) in order to reproduce the observed radial velocity dispersion profiles.
%(most recently literature on model calculations of 47 Tuc velocity dispersion profiles can be found in Ref. \cite{Mann2018}).

In this work, we use these conclusions to show that the observations of velocity dispersion profiles from dwarf galaxies have severely 
constrained the abundance of PBHs in DM since the presence of PBHs in dwarf galaxies
will lead to a large velocity dispersion anisotropy as that shown in the globular clusters.
As an example, we use the dwarf galaxy Leo I, which has an age of $\sim (7-10)$ Gyr and is located at a distance 
about 250 kpc from the Sun.
The half-mass relaxation time of the cluster can be computed as \cite{Spitzer1987}
\begin{eqnarray}
    \tau_{\rm rh}=\frac{1.7\times 10^{5}~(r_{\rm h}/{\rm pc})^{3/2}~N^{1/2}}{(m_{\rm s}/M_{\odot})^{1/2}}~{\rm yr},
\end{eqnarray}
where $r_{\rm h}$ is the half-light radius and $N$ is the total number of stars.
For Leo I, the magnitudes of the half-light radius and total mass are about 
$100~{\rm pc}$ and $10^6~M_{\odot}$ \cite{McConnachie2012}.
%$\sim \mathcal{O}(100)~({\rm pc})$ and $\sim \mathcal{O}(10^6)~({M_{\odot}}$
Thus, the ratio of stellar age to half-mass relaxation time for Leo I is estimated to be $t_{\rm age}/\tau_{\rm rh}\sim 0.1-1$.
Other dwarf galaxies with similar relaxation times include Sculptor and Carina.
In this stage, the cluster has large radial velocity dispersion if the heavy 
components are present in the system \cite{Peuten2017}.

In Sec. II of this work, we introduce the Michie–King model for the description of stellar radial velocity dispersion profiles
when PBHs are present in the cluster system. We present our data analysis and the results in Sec. III. Finally, we give a summary
of our conclusions in Sec. IV.

\section{The radial velocity dispersion model}

One approach to model the radial velocity dispersion profiles with the presence of PBHs is by solving the 
Jeans' equation, like that done in Refs. \cite{Meylan1986AA, Mann2018}.
Another approach to describe the kinematical anisotropy is via the Michie-King model \cite{Michie1963}, 
for instance, see Refs. \cite{Peuten2017, Gieles2015, Meylan1987AA, Gunn1979AJ}.
The Michie-King model for the phase-space density of mass component $j$ is given by
\begin{eqnarray}
    f_{{\rm MK},j}(E,L)=\exp\left\{ -L^{2}/(2r_{{\rm a},j}^2\sigma_{j}^2) \right\}f_{{\rm K},j}(E),
\end{eqnarray}
where $j=(\rm s,PBH )$ represents the component of star and PBH, 
the angular momentum $L=vr\sin\theta$ and energy $E=v^2/2+\Phi(r)$, where $\Phi(r)$ is the gravitational potential.
$\sigma_{j}$ is the velocity dispersion of the component $j$ at the center. 
The King phase-space distribution function \cite{King1965, King1966} of mass component $j$ is given by
\begin{eqnarray}
    f_{{\rm K},j}(E)=\frac{n_{0,j}}{(2\pi\sigma_{j}^2)^{3/2}}\left\{ \exp\left( \frac{\Phi(r_{\rm t})-E}{\sigma_{j}^2} \right)-1 \right\},
\end{eqnarray}
where $n_{0,j}$ is a normalized constant,
$r_{\rm t}$ is the tidal radius which represents the external edge of the system.
The truncation in energy is introduced to mimic the role of tides of the galaxy. 
The King model describes a cluster with an isotropic velocity dispersion profile, 
it is thought to be a correct zeroth-order dynamical reference model to represent quasirelaxed stellar systems.
The Michie-King model is a natural extension of the King model to the anisotropic case.

One of the significant parameters in the Michie-King distribution function is the anisotropic radius of the mass component $j$, $r_{{\rm a},j}$.
Inside the anisotropic radius, the velocity dispersion is nearly isotropic, while outside the anisotropic radius 
the anisotropy becomes significant. Following Refs. \cite{Gieles2015, Peuten2017}, the anisotropic radius of 
component $j$ has a power-law dependence on the mass $m_{j}$,
\begin{eqnarray}
    r_{{\rm a},j}=r_{\rm a}\mu_{j}^{\eta},
\end{eqnarray}
where $r_{\rm a}$ is the anisotropic radius of the system, $\eta\simeq 0.5$ for Leo I \cite{Peuten2017}, and $\mu_{j}=m_{j}/\bar{m}$, 
here $\bar{m}\simeq m_{\rm PBH}$
%$\bar{m}=\sum_{j}m_{j}\rho_{0,j}/\sum_{j}\rho_{0,j}$ 
is the mean mass at the cluster center \cite{Gunn1979AJ}.
%$\rho_{0,j}$ is the central density of mass component $j$. 
The anisotropic radius of the system $r_{\rm a}$ is related to the cluster core radius 
$r_{\rm c}$ by the relation $r_{\rm a}=xr_{\rm c}$, where the constant $x\lesssim 10$ \cite{Peuten2017, Meylan1987AA}.
When many components appear in the system, the central density rapidly becomes dominated by the most massive 
component and the core radius quickly becomes smaller, exhibiting a much deeper core collapse \cite{Giersz1996}.
Thus, when PBHs are contained in the system, the core radius of the cluster can be estimated as
\begin{eqnarray}
    r_{\rm c}^2\simeq\frac{9\sigma_{\rm PBH}^2}{4\pi G\rho_{0,\rm PBH}}.
\end{eqnarray}
where $\sigma_{\rm PBH}$ and $\rho_{0,\rm PBH}$ are the velocity dispersion and mass density of PBH at the center of the cluster.
The velocity dispersion of star and PBH at the center of the cluster are related by the equipartition
\begin{eqnarray}
    m_{\rm PBH}\sigma_{\rm PBH}^2=m_{\rm s}\sigma_{\rm s}^2,
\end{eqnarray}
We only consider the PBHs at mass scale $m_{\rm PBH}\ge 5~M_{\odot}$.
The central density of PBHs is related to DM density at the center by $\rho_{0,\rm PBH}=f_{\rm PBH}\rho_{0,\rm DM}$,
where $f_{\rm PBH}$ is the abundance of PBHs in DM and $\rho_{0,\rm DM}$ is the DM central density.

Although under the framework of $\Lambda CDM$ cosmology, large N-body numerical simulations lead to the commonly used 
Navarro-Frenk-White (NFW) halo cusp spatial density profile, analysis of observations in the central regions of various 
dwarf halos is in favor of cored profiles (a review on this topic can be found in Ref. \cite{Tulin2017PR}).
Thus, for the DM mass density profile of the dwarf galaxy, we adopt the Burkert model \cite{Burkert1995}
\begin{eqnarray}
    \rho_{\rm DM}=\frac{\rho_{\rm b}}{(1+x_{\rm b})(1+x_{\rm b}^2)},
\end{eqnarray}
where $x_{\rm b}=r/r_{\rm b}$. This model has two parameters, the scale radius $r_{\rm b}$ and the central density of 
DM $\rho_{\rm b}$. The Burkert model describes a cored density profile.
The corresponding gravitational potential is
\begin{eqnarray}\label{gp}
    \frac{\Phi(r)}{\Phi_{\rm b}}&=&\left ( 1-\frac{1}{x_{\rm b}}\right )\frac{\ln(x_{\rm b}^2+1)}{4} \nonumber\\
&+&\left ( 1+\frac{1}{x_{\rm b}}\right )\left ( \frac{\tan^{-1}(x_{\rm b})}{2}-\frac{\ln(x_{\rm b}+1)}{2} \right ),
\end{eqnarray}   
where $\Phi_{\rm b}=4\pi G\rho_{\rm b}r_{\rm b}^2$. Here we have defined the potential to be 0 at the center of the system
and to be $\pi \Phi_{\rm b}/4$ at infinity \cite{Strigari2017}. For the calculations of the distribution and velocity dispersion of the stars,
we are only concerned with the relative gravitational potential $\Psi(r)=\Phi(r_{\rm t})-\Phi(r)$.
%only the relative gravitational potential $\Psi(r)=\Phi(r_{\rm t})-\Phi(r)$ is concerned.
Since DM is the major component of dwarf galaxies, we treat Eq. (\ref{gp}) as the total gravitational potential.
%--------------------------------------------------------
\begin{figure}[ht]
    \includegraphics[width=85mm,angle=0]{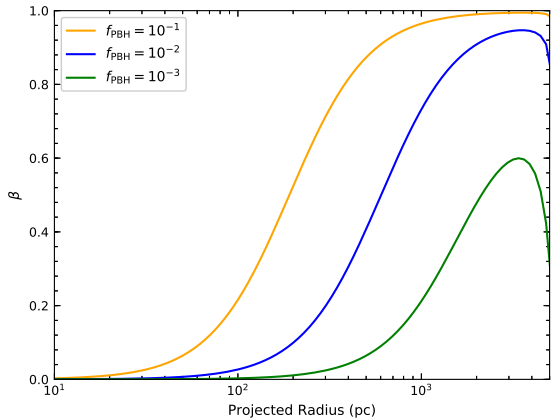}
    \caption{Anisotropy of stellar velocity dispersion as a function of the projected radius.
    The yellow, blue, and green lines correspond to the anisotropy with the values of 
    $f_{\rm PBH}=10^{-1}$, $10^{-2}$, and $10^{-3}$, assuming $m_{\rm PBH}=30~M_{\odot}$.}
    \label{bt}
\end{figure}
%---------------------------------------------------------

For a given phase-space distribution function $f_{j}(r,v)$, the mass density profile of mass component $j$ is given by 
\begin{eqnarray}
    \rho _{j}(r)=2\pi\int_{0}^{\pi}d\theta \sin\theta \int _{0}^{v_{\rm e}}dvv^{2}f_{j}(r,v),
\end{eqnarray}
where $v_{\rm e}(r)=\sqrt{2\Psi(r)}$ is the escape velocity. For the King model we have
\begin{eqnarray}
    \rho _{{\rm K},j}(r)&=&n_{0,j}\Bigg\{ \exp\left ( \frac{\Psi}{\sigma_{j}^2} \right ){\rm Erf}
    \left ( \frac{\sqrt{\Psi}}{\sigma_{j}} \right )\nonumber\\
    &-&\sqrt{\frac{4\Psi}{\pi\sigma_{j}^2}}\left ( 1+\frac{2\Psi}{3\sigma_{j}^2} \right ) \Bigg\}.
\end{eqnarray}

The radial and tangential stellar velocity dispersion profiles are given by
\begin{eqnarray}
    \rho _{j}\sigma_{r}^2(r)=2\pi\int_{0}^{\pi}d\theta\cos^2\theta  \sin\theta \int _{0}^{v_{\rm e}}dvv^{4}f_{j}(r,v),\\
    \rho _{j}\sigma_{\theta}^2(r)=\pi\int_{0}^{\pi}d\theta\sin^2\theta  \sin\theta \int _{0}^{v_{\rm e}}dvv^{4}f_{j}(r,v).
\end{eqnarray}
Obviously, radial and tangential velocity dispersion of mass component $j$ are equal in the King model, 
i.e., $\sigma_{r}^2=\sigma_{\theta}^2=\sigma_{\rm K}^2$. Again we have
\begin{eqnarray}
    \rho _{{\rm K},j}\sigma_{\rm K}^2(r)&=&n_{0,j}\Bigg\{  \sigma_{j}^2\exp\left ( \frac{\Psi}{\sigma_{j}^2} \right )
    {{\rm Erf}\left ( \frac{\sqrt{\Psi}}{\sigma_{j}} \right )}\nonumber\\
    &-&\frac{2\sqrt{\Psi}}{15\sqrt{\pi}\sigma_{j}^3}\left ( 15\sigma_{j}^4+10\sigma_{j}^2\Psi+4\Psi^2 \right ) \Bigg\}.
\end{eqnarray}
The projected stellar density profile and line-of-sight velocity dispersion profile are given by
\begin{eqnarray}
    I_{\rm s}(R)&=&2\int _{0}^{\infty }\rho _{\rm s}(r)dz,\\
    I_{\rm s}\sigma_{\rm los}^2(R)&=&2\int _{0}^{\infty }\rho _{\rm s}(r)\frac{z^2\sigma_{r}^2+R^2\sigma_{\theta}^2}{z^2+R^2}dz,
\end{eqnarray}
where $R=\sqrt{r^2-z^2}$ is the projected distance.
The anisotropy of velocity dispersion is defined as
\begin{eqnarray}
    \beta(r)=1-\sigma_{\theta}^2(r)/\sigma_{r}^2(r).
\end{eqnarray}
In Fig. \ref{bt} we plot the stellar anisotropy with various values of the PBH abundance $f_{\rm PBH}$.
We have fixed the PBH mass $m_{\rm PBH}=30~M_{\odot}$; the other parameters used here can be found in Table \ref{tab}.
The stellar anisotropic radius $r_{\rm a,s}$ (in unit pc) can be estimated by
\begin{eqnarray}
    r_{\rm a,s}\simeq 129\left ( \frac{\sigma_{\rm s}}{{~\rm km/s}} \right )\left ( \frac{f_{\rm PBH}\rho_{\rm b}}
    {M_{\odot}/{\rm pc^{3}}}\right )^{-\frac{1}{2}}\left ( \frac{m_{\rm PBH}}{M_{\odot}} \right )^{-1}.
\end{eqnarray}
For the value of $f_{\rm PBH}=10^{-2}$, the stellar anisotropic radius $r_{\rm a,s}\simeq 600~\rm pc$.
There are two opposite physical regimes in the Michie–King models \cite{Amorisco2012}. 
In the isothermal region, $\Psi(r)/\sigma_{\rm s}^2\gg 1$ (the King density profile is controlled by the exponential term),
the anisotropy profile can be simplified as $\beta\sim r^2/(r^2+r_{\rm a,s}^2)$.
As expected, the system is isotropic at the center and it radially becomes anisotropic at radius $r>r_{\rm a,s}$.
In the regions near the tidal radius, the potential $\Psi(r)$ tend to 0, and the anisotropy asymptotes to
$\beta\sim 2r^2\Psi(r)/(9r_{\rm a,s}^2\sigma_{\rm s}^2)$. As shown in the figure, the anisotropy
radially decreases to 0 in these regimes.
%The stability of the stellar radial orbit inside the tidal radius require $\beta(r)<1/2$ for $r<r_{\rm t}$, which
%puts the constraints $f_{\rm PBH}\lesssim 10^{-3}$ for $m_{\rm PBH}=30~M_{\odot}$.
%--------------------------------------------------------------------------
\begin{figure}[ht]
    \includegraphics[width=85mm,angle=0]{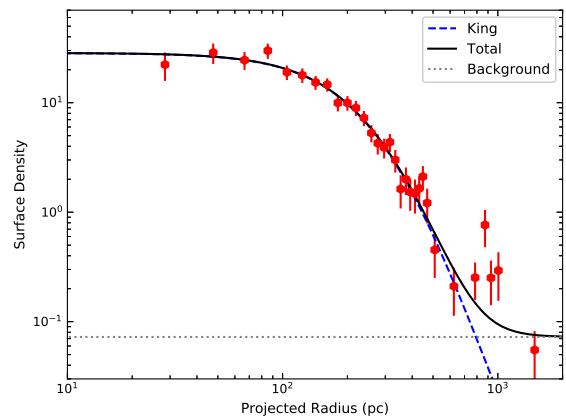}
    \caption{Stellar surface density as a function of the projected radius. The dashed blue line shows the results of the King model, 
    the dotted grey line corresponds to the background, and the black line represents the total surface density of the model.}
    \label{sd}
\end{figure}
%--------------------------------------------------------------------------

\section{Data analysis}

First, we use the ``null'' model, i.e., the isotropic King model [which corresponds to $r_{\rm a}\to \infty$
($f_{\rm PBH}=0$) in the Michie-King model] to fit the surface density \cite{Irwin1995MNRAS} and velocity 
dispersion \cite{Walker2007} observations from Leo I.
In this framework, the distributions of DM and the stars are simultaneously determined.
We use the Markov chain Monte Carlo (MCMC) method to derive the posterior probability distributions of the parameters 
from the observational data \cite{Lewis1995, Putze2009}. 
Such an approach is based on the Bayesian theorem, in which the posterior probability distributions 
of the parameters are directly linked to the likelihood function
\begin{eqnarray}
    \mathcal{L}(\vartheta )\propto \exp(\chi^2(\vartheta )/2),
\end{eqnarray}
where $\vartheta=\{\rho_{\rm b},r_{\rm b},r_{\rm t}, \sigma_{\rm s},n_{0,\rm s},b_{0}\}$ is the set of free parameters 
in the King model ($b_{0}$ is the stellar background surface density), the chi-square
$\chi^2(\vartheta )=\sum_{i}^{n}(\mu_{i}^{\rm d} -\mu_{i}^{\rm m} )^2/\Delta_{i}^2$, where $\mu_{i}^{\rm d}$ is the data,
$\mu_{i}^{\rm m}$ is the value predicted by the model, and $\Delta_{i}$ is the deviation of the measurements.
%*****************************Fig.1***************************************

\begin{figure}[ht]
    \includegraphics[width=85mm,angle=0]{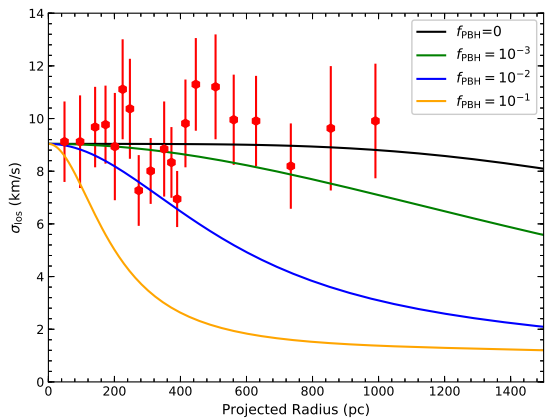}
    \caption{Stellar velocity dispersion as a function of the projected radius. The black line is the results given by the King model
    (or the Michie-King model with $f_{\rm PBH}=0$). The yellow, blue, and green lines correspond to the velocity dispersion predicted 
    by the Michie-King models with the values of $f_{\rm PBH}=10^{-1}$, $10^{-2}$, and $10^{-3}$, assuming $m_{\rm PBH}=30~M_{\odot}$.}
    \label{vd}
\end{figure}

%------------------------------------------------------------------------------

The best-fit set of parameters corresponds to a minimum value of $\chi_{\rm tot}^2=\chi_{\rm sd}^{2}+\chi_{\rm vd}^{2}$, 
where $\chi_{\rm sd}^{2}$ and $\chi_{\rm vd}^{2}$ are the chi-square of surface density and velocity dispersion.
To ensure our $\chi_{\rm tot}^{2,\rm min}$ is robust, we have run several chains from different starting points.
Figures \ref{sd} and \ref{vd} (black line) demonstrate that the observations are consistent with the predictions of the 
isotropic King model, which is confirmed by the minimum values of chi-square for the fits: $\chi_{\rm sd}^{2,\rm min}=25.0$ for the 
photometry (31 data points) and $\chi_{\rm vd}^{2,\rm min}=12.9$ for the 
kinematics (20 data points). The fitting results are shown in Table \ref{tab}.
In the depictions of Figs. \ref{sd} and \ref{vd}, we have fixed the parameters in the set $\vartheta$ at the best-fit values.

%*************************************************************************
\begin{table}[htbp]
    \centering
    \caption{\label{tab} List of fitting results with the King model}
    \begin{threeparttable}
    \begin{tabular}{cccccccccccccc}
    \toprule
    \hline
    \hline
    & Parameter & Best fit & Mean & 95\% credible intervals \\
    \hline
    & $r_{\rm t}$~(pc) & 5419.50 & 5291.15 & [4573.83,~6489.03] \\
    & $r_{\rm b}$~(pc) & 318.91 & 319.57 & [295.32,~343.93] \\
    & $\rho_{\rm b}$~($10^{-2}~M_{\odot}/{\rm pc}^3$) & 4.60 & 4.47 & [3.95,~5.11] \\
    & $\sigma_{\rm s}$~(km/s) & 3.00 &2.96 & [2.92,~3.02] \\
    \midrule    
    \hline
    \hline
    \bottomrule
    \end{tabular}
    %\begin{tablenotes}
    %\footnotesize
    % \item[\it a] $\alpha_{X}$ is fixed at $0.1$.
    % \item[\it b] $J_S(\varepsilon_{\phi })$ and $J_{\rm obs}$ in unit $\rm GeV^{2}\;cm^{-5}$.
    %\end{tablenotes}
    \end{threeparttable}
\end{table}
%-----------------------------------------------------------------------------------------------------------------

In Fig. \ref{vd} we show the stellar velocity dispersion predicted by the Michie-King model with various values of $f_{\rm PBH}$.
The kinematical data points depicted in the figure keep nearly a constant up to a radius $\sim 1000$ pc.
When the PBHs are present, the radial velocity dispersion develops and exceeds the tangential one at large radii;
In the extreme case, a deep core collapse may lead to the formation of a massive black hole \cite{Gurkan2004ApJ}, 
a steep rise in kinematical data near the center may indicate such a massive 
black hole inhabiting the center of the cluster \cite{Binney1982}.
%this leads the depression of the line-of-sight velocity dispersion profile.
%--------------------------------------------------------------------------
\begin{figure}[ht]
    \includegraphics[width=85mm,angle=0]{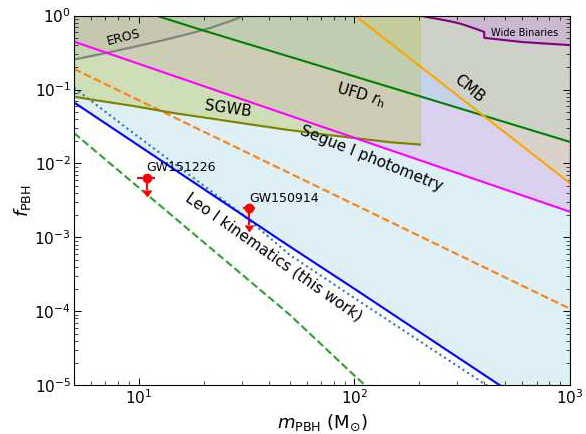}
    \caption{Constraints on the PBH abundance $f_{\rm PBH}$ as a function of PBH mass $m_{\rm PBH}$,
    the colored regions are excluded by the constraints. Our constraints from Leo I kinematics are shown by the solid blue line.
    Previous constraints on the PBH abundance, including constraints from the microlensing experiment EROS \cite{EROS2007}, 
    from stochastic GW background (SGWB) \cite{Wang2017PRL}, from wide binaries\cite{Monroy2014}, from cosmic microwave 
    background (CMB) photoionization \cite{Ali2017}, from the half-light radius of the ultrafaint dwarfs (UFDs) \cite{Brandt2016}, 
    and from the surface density of Segue I \cite{Koushiappas2017} are also shown in the figure. 
    The two data points \cite{Wang2017PRL} represent the required PBH abundance for the explanations of 
    GW150914 \cite{GW150914} and GW151226 \cite{GW151226}.}
    \label{lt}
\end{figure}
%--------------------------------------------------------------------------

We use the observations from Leo I to constrain the anisotropy of velocity dispersion when the PBHs are present in the DM.
For each given value of $m_{\rm PBH}$, the density and line-of-sight velocity dispersion profiles are computed 
using the Michie-King model and compared with the observed data.
We obtain the posterior probability distribution $\mathcal{P}(f_{\rm PBH},\vartheta)|_{m_{\rm PBH}}$ 
for a given value of PBH mass by the MCMC analysis.
Using a procedure called marginalization \cite{Cowan1997}, the information about the parameter of PBH abundance is extracted by integrating 
over all other parameters $\vartheta$ in the posterior density.
This yields the posterior probability distribution of PBH abundance $\mathcal{P}(f_{\rm PBH})|_{m_{\rm PBH}}$ for a given value of PBH mass.
Finally, by integrating the posterior probability distribution of PBH abundance to a certain value $f_{\rm up}$ so that 
$\int_{0}^{f_{\rm up}}\mathcal{P}(f_{\rm PBH})|_{m_{\rm PBH}}df_{\rm PBH}=0.9999$,
we are able to determine the upper limit on the PBH abundance $f_{\rm PBH}$ at a 99.99\% confidence level.

Our constraints on the PBH abundance $f_{\rm PBH}$ as a function of PBH mass are shown by the blue line in Fig. \ref{lt},
the colored regions are excluded by the constraints.
The resulting relation between $f_{\rm PBH}$ and $m_{\rm PBH}$ represented by the blue line is given by
\begin{eqnarray}\label{f}
    f_{\rm PBH}\simeq 2.0\times(1~M_{\odot}/m_{\rm PBH})^2.
\end{eqnarray}
For a PBH with mass $30~M_{\odot}$, the constraint on the PBH fraction $f_{\rm PBH}$ is up to $\sim 2\times 10^{-3}$.
One of the uncertainties in deriving our constraints may contribute from the parameter $\eta$, which is determined by the 
dynamical stage of the cluster. The limits on the PBH abundance in DM assuming the value of 0.2 and 0.8 for $\eta$ are shown by
the dashed yellow and dashed green lines in Fig. \ref{lt}. The results show that the constraints can be altered at most by about
one order of magnitude when $\eta$ ranges from 0.2 to 0.8. 
Moreover, we have used the Burkert profile as the underlying DM distribution. The limits on PBHs assuming NFW 
DM halo \cite{NFW1996} are also shown in the same figure (the dotted blue line). For $m_{\rm PBH}\gtrsim 20 M_{\odot}$ the constraints on the PBHs 
abundance with the NFW profile are a little stronger than that of using Burkert profile, while for light PBH masses, the limits become weaker.

Previous constraints on the PBH abundance, including constraints from the microlensing experiment EROS \cite{EROS2007}, from SGWB 
\cite{Wang2017PRL}, from wide binaries\cite{Monroy2014}, from CMB photoionization \cite{Ali2017}, 
from the half-light radius of the UFDs \cite{Brandt2016}, and from the surface density of Segue I \cite{Koushiappas2017} 
are also shown in the figure. The two data points \cite{Wang2017PRL} represent the required PBH abundance for the explanations of 
GW150914 \cite{GW150914} and GW151226 \cite{GW151226}.
Since the distributions of PBHs in the early Universe are more dense, a fraction $10^{-2}-10^{-3}$ of the DM
consisting of PBHs can explain the merger rate estimated by LIGO Collaboration \cite{Sasaki2016, Wang2017PRL, Raidal2018}. 
As shown in the figure, our limits have already extended into these parameters regimes; thus, we have the confidence to conclude that the 
primordial origin of the GW events observed by the LIGO is stringently constrained by the kinematical observation from dwarf galaxies.

\section{Conclusions and outlook}

In summary, we have proposed that the observations of the velocity dispersion profile from the dwarf galaxies can be used to constrain the 
PBH abundance in the dark matter.
The observed increase of velocity dispersion near the center of the globular cluster $\omega$ Cen and 47 Tuc indicates that 
a fraction $1\%-10\%$ of the total masses should be in the heavy dark remains with masses $\sim 2~M_{\odot}$ \cite{Meylan1986AA}.
The same effects on the velocity dispersion profiles are expected when the PBHs are present in the dwarf galaxies,
where dark matter is believed to be the dominant mass component.
We use the kinematical observations from Leo I as an example to demonstrate that the PBH fraction 
$f_{\rm PBH}\gtrsim 2.0\times(1~M_{\odot}/m_{\rm PBH})^2$ is ruled out at 99.99\% confidence level.
These give the most stringent limits on the PBH abundance at the mass scales $\sim(1-10^3)~M_{\odot}$ 
and tightly constrain the primordial origin of GW events observed by the LIGO experiments.

We have assigned a single stellar mass (1~$M_{\odot}$) for the star cluster,
further improvement of our analysis could be done by taking into account the stellar mass function in the cluster.
This would involve a small fraction of dark remains, for instance, the white dwarfs, neutron stars, and stellar black holes.
%Our constraints would be stronger if we use the Navarro-Frenk-White (NFW) DM halo \cite{NFW1996}, which has a cusp spatial density profile.
The future kinematical observations from the dwarf galaxies could improve both the estimation of the dwarf galaxies' DM content 
and the constraints on the PBH abundance in DM.

{\it Acknowledgments} 
B. Q. L. thanks Misao Sasaki and Mohammadreza Zakeri for useful discussions and comments on this manuscript. 
This work is supported by the NSFC under Grants No. 11851302, No. 11851303, No. 11747601, 
No. 11690022, and the Strategic Priority Research Program of the Chinese Academy of Sciences with Grant No. XDB23030100 
as well as the CAS Center for Excellence in Particle Physics (CCEPP). This work is also supported in part by the MOST Grant 
No. 108-2811-M-002-524.

%\appendix

\end{document}